\documentclass[12pt]{report}
\usepackage{amsmath,amsthm,amssymb}
\usepackage{mathrsfs}
\usepackage{hyperref}
\usepackage[a4paper]{geometry}
\usepackage{fancyhdr}
\usepackage{lastpage}
\usepackage{graphicx, wrapfig, subcaption, setspace, booktabs}
\usepackage[english]{babel}

\hypersetup{
    colorlinks=true, 
    linktoc=all,     
    linkcolor=black,  
}
\theoremstyle{plain}
\newtheorem{thm}{Theorem}
\newtheorem{defn}{Definition}
\newtheorem{lem}{Lemma}
\newtheorem{prop}{Proposition}
\newtheorem{alg}{Algorithm}
\newtheorem*{cor}{Corollary}
\newcommand{\HRule}[1]{\rule{\linewidth}{#1}}
\begin{document}

\title{ \normalsize \textsc{}
		\\ [-4.0cm]
		\HRule{0.5pt} \\
		\LARGE \textbf{\uppercase{Codes that achieve capacity on symmetric channels}\footnote{ \textit{ As part of the Visiting Students Research Programme at the School of Technology and Computer Science, Tata Institute of Fundamental Research, Mumbai, India -40005.}}}
		\HRule{2pt} \\ [0.5cm]
		\normalsize \today \vspace*{5\baselineskip}}
\date{}
\author{
		Vishvajeet Nagargoje\footnote{Email : \href{mailto :nvishvajeet@gmail.com}{nvishvajeet@gmail.com}, Contact : +91-9884299504. }\\ 
Indian Institute of Technology Madras\\ \vspace {5 mm}
Under the guidance of\\ \vspace{5 mm} Prof. Prahladh Harsha\footnote{Email: \href{mailto:prahladh@tifr.res.in}{prahladh@tifr.res.in}}\\ Tata Institute of Fundamental Research, Mumbai}

\maketitle

\newpage
\begin{center}\begin{large}{ACKNOWLEDGEMENT}\end{large}\end{center}

I would like to express my deep gratitude to Prof. Prahladh Harsha and Prof. Vinod Prabhakaran, my internship supervisors, for giving me an opportunity to work under them during the summer. I would also like to thank Sasank Mouli, for helping me out  with things.

\vspace{2mm}
I would also like to acknowledge Tata Institute of Fundamental Research, Mumbai for warmly hosting me in the summer.

\vspace{2mm}

\begin{abstract}

Transmission of information reliably and efficiently across channels is one of the fundamental goals of coding and information theory. In this respect, efficiently decodable deterministic coding schemes which achieve capacity provably have been elusive until as recent as 2008, even though schemes which come close to it in practice existed. This survey tries to give the interested reader an overview of the area.\\

Erdal Arikan came up with his landmark polar coding shemes which achieve capacity on symmetric channels subject to the constraint that the input codewords are equiprobable. His idea is to convert any B-DMC into efficiently encodable-decodable channels which have rates 0 and 1, while conserving capacity in this transformation. An exponentially decreasing probability of error which independent of code rate is achieved for all rates lesser than the symmetric capacity. These codes perform well in practice since encoding and decoding complexity is $O(N \log N).$ Guruswami et al. improved the above results by showing that error probability can be made to decrease doubly exponentially in the block length.\\

We also study recent results by Urbanke et al. which show that 2-transitive codes also achieve capacity on erasure channels under MAP decoding. Urbanke and his group use complexity theoretic results in boolean function analysis to prove that EXIT functions, which capture the error probability, have a sharp threshold at 1-R, thus proving that capacity is achieved. One of the oldest and most widely used codes - Reed Muller codes are 2-transitive. Polar codes are 2-transitive too and we thus have a different proof of the fact that they achieve capacity, though the rate of polarization would be better in Guruswami's paper.

\end{abstract}
\tableofcontents


\newpage
\section*{Introduction}
\addcontentsline{toc}{section}{Introduction}
Communication is a fundamental need of our lives. In modern times even though the need for communication and the (ever-the-more sophisticated) tools available with us have increased, communication is something which we humans have been doing for a long time. Inherently, we have error correcting capabilities built into us, which help us understand our fellow humans even when there is corruption due to say, speech defects, physical mediums, etc. We want our communication systems to do the same for us, but unfortunately there are fundamental questions which we have not been able to answer, and things don't look that easy.\\

We, as people studying computer science like to abstract things out and work with appropriate models. In this pursuit, Alice and Bob come to our rescue, as always.

Suppose Alice and Bob want to communicate with each efficiently other over a 'noisy' physical transmission medium which corrupts the information. We abstract out the possible scheme of corruptions into something which we shall call a $'channel'$.

\begin{defn}{Communication Channel}

\noindent We define a discrete channel to be a system consisting of the following : \begin{itemize} \item An input alphabet $\chi$
\item An output alphabet $\Upsilon$ 
\item A conditional probability transition matrix $P(y|x)$
\end{itemize}
\end{defn}

\noindent  The conditional probability distribution expresses the probability of observing the output symbol y given that we send the symbol x. The channel is said to be \textit{ memoryless} if the probability distribution of the output depends only on the input at that particular time and is independent of the previous channel inputs or outputs. We would be talking about binary memoryless channels in the rest of this article.

Let's quickly define the information theoretic functions that we would be using in the course of this article.

\begin{defn}{Entropy of a random variable}

The entropy H(X) of a discrete random variable X is defined by 
\center $H(X) = - \underset{x \in \chi}{\sum} p(x) \log p(x)$.
\end{defn}
\begin{defn}{Mutual information between two random variables}

 The mutual information I(X;Y) between two random variables X and Y is defined to be 

\center $I(X;Y) = \underset {x \in \chi}{\sum}\underset{y \in \Upsilon}{\sum} p(x,y) \log \frac{p(x,y)}{p(x)p(y)}$
\end{defn}
\begin{defn}{(Shannon)Capacity of Channel}

 The "information" channel capacity of a discrete memoryless channel is defined as 

\center C = Max$_{\text P(x)}$ I (X;Y)
\end{defn}

Observe that the capacity of a channel is completely characterized once we specify the input output conditional probability distribution.

Let us look at a few frequently occuring discrete channels.
\begin{defn}{ Noiseless Binary Channel }

Also called 'perfect channel', this is a channel such that : \begin{itemize}
\item $\chi = \{ 0, 1\}$
\item $\Upsilon = \{ 0, 1\}$
\item P( y = 0 | x = 0) = 1 and P( y = 1 | x = 1) = 1
\end{itemize}
\end{defn}

Observe that we do not need any coding scheme if we use the noiseless channel. 
\begin{defn}{Binary Erasure Channel}

An erasure channel with erasure probability p has the following parameters : \begin{itemize}
\item $\chi = \{ 0, 1\}$
\item $\Upsilon = \{ 0, 1, e\}$
\item P( y = 0 | x = 0) = 1 - p , P( y = 1 | x = 1) = 1 - p\\
P( y = e | x = 1) = p and P( y = e | x = 1) = p
\end{itemize}
\end{defn}

\begin{defn}{Binary Symmetric Channel}

A binary symmetric channel with flip-over probability p has the following parameters : \begin{itemize}
\item $\chi = \{ 0, 1\}$
\item $\Upsilon = \{ 0, 1\}$
\item P( y = 0 | x = 0) = 1 - p , P( y = 1 | x = 1) = 1 - p\\
P( y = 1 | x = 0) = p and P( y = 0 | x = 1) = p
\end{itemize}
\end{defn}

\begin{defn}{Symmetric Channel}

A channel for which there exists a permutation $\pi$ of the output alphabet $\Upsilon$ such that :\begin{itemize}

\item $\pi^{-1} = \pi $
\item $ W(y|1) = W(\pi (y)|0)$  $\hspace{1.5mm}$ $\forall y \in \Upsilon$
\end{itemize}
Observe that the symmetric capacity I(W) equals the Shannon capacity when W is a symmetric channel. We can also see that the BEC and the BSC are symmetric channels.

\end{defn}

Observe that in all channels except the noiseless channel we cannot decode correctly unless we use encoding and decoding schemes. Intuitively it is clear that if we add some amount of redundancy in the code, it would be easier for us to correct errors. But this leads to transmission of more bits than the message actually contains. The fundamental question in Information and Coding Theory is the tradeoff between redundancy and the number of errors that can be corrected. We shall formalize the notion of redundancy in a code.
\begin{defn}{Rate of Code}

The rate of a code is defined to be equal to $\frac{\text{dimension}}{\text{block length}}$

\end{defn}
 Claude Shannon gave an operational definition of the channel capacity, which implies that it is the maximum rate at which we can transmit information across a channel reliably, with error going to zero in the limit of the block length going to infinity.
\begin{thm}{Shannon's coding theorem}

Given a noisy channel with channel capacity C and information transmission rate R, then if R < C there exist codes that allow the probability of error at the receiver to be made arbitrarily small, and the converse is also true.

\end{thm}
Shannon's used a \textit{random} coding approach in his landmark paper. But this random coding approach is not something which satisfies theoretical computer science people like us and we want to \textit{lay our hands} on a particular coding scheme which does the above for us reliably. We also want to be able to decode it efficiently. In other words, we want \textit{an efficiently decodable capacity achieving deterministic coding scheme}.\\

We would like to analyze how bad our coding scheme is, by looking at the error probability when a trivial decoding scheme is used - ML / MAP decoding, which looks at the codeword closest to the one received and outputs it. In doing so we would require a parameter defined as follows :

\begin{defn}{Bhattacharya parameter}

For a channel W Bhattacharya parameter is defined as, 
\center Z(W) $\triangleq  \underset{y \in \Upsilon}{\sum}\sqrt{W(y|0)W(y|1)}$.
\end{defn}

\begin{thm}{The Bhattacharya parameter is an upper bound on the probability of error achieved by ML/MAP decoding.}

\end{thm}

We can quickly relate the symmetric capacity I(W) and Z(W), the Bhattacharya parameter. Intuitively one would expect that I(W) $\approx$ 0 if Z(W) $\approx$ 1. The following bounds make this precise : 

\begin{prop} Relation between symmetric capacity and Bhattacharya parameter.

\center I(W) $\geq \log {\frac 2 {1+Z(W)}}$
\center I(W) $\leq \sqrt {1- Z(W)^2}$
\end{prop}

\newpage
\section*{Polar Codes}
\addcontentsline{toc}{section}{Polar Codes}
A deterministic capacity achieving coding scheme has been elusive for a long time, and in the process computer scientists have come up with different types of schemes which come extremely close to achieving capacity in practice but fail to do so provably. It was in 2008 that Erdal Arikan came up with his landmark coding scheme which achieves capacity on symmetric channels. His work is a culmination of more than 20 years of reserach into the sequential cutoff rate for different types of codes. This section tries to outline the idea behind Arikan's \textit{polar codes}.

\subsection*{Channel Polarization}
\addcontentsline{toc}{subsection}{Channel Polarization}
It can be observed that it very easy to code two types of channels - the perfect and the useless channel. The main idea behind Arikan's technique is that it manufactures out of N independent copies of a given BDM\footnote{Binary Discrete Memoryless Channel} W, a second set of channels which is polarized i.e. consisting of only perfect and useless (a channel in which the output is independent of the input) channels and this transformation also conserves capacity. This operation goes through two parts : combining and splitting. Let's look at these operations in detail.

\begin{itemize}

\item \textbf{Channel combining}

We take N =$2^n$ channels, each denoted by W, and manufacture a vector channel $W_N : \chi^N \longrightarrow \Upsilon^N$ recursively.

In general, 2 copies of $W_{\frac N2}$ are combined to produce $W_N$ so that the the input $u_1^N$ to $W_N$ is first transformed to $s_1^N$ where \begin{center} $s_{2i-1}=u_{2i-1} \oplus u_{2i} $ and $s_{2i}=u_{2i}$ for $ 1 \leq i \leq \frac N2$
\end{center}

For the 0-th level of recursion (n=0) we set $W_1 \triangleq W$

The first level of recursion combines to copies of $W $ and obtains the channel $W_2$ where $W_2(y_1,y_2|u_1,u_2) = W (y_1|u_1\oplus u_2) W(y_2|u_2)$ as and so on.

The operator $R_N$ is a permutation called the \textit{reverse shuffle operation} and acts on the input $s_1^N$ to produce $v_1^N = ( s_1, s_3, s_{N-1}, \dots s_2, s_4 \dots s_N)$, which is the input to the 2 copies of $W_\frac N2$.

At each level of recursion, it should be observed that the mapping from $u_1^N \longrightarrow v_1^N$ is linear over $GF(2)$. Inductively, it can be proved that the  mapping from $u_1^N \longrightarrow x_1^N$ which takes the input from $W_N$ to $W^N$ channel is also linear and can be denoted by $x_1^N = u_1^N G_N$.
 
We can thus relate the transition probabilities of the 2 channels $W_N$ and $W^N$ by the following equation : 

\begin{center}$W_N(y_1^N | u_1^N) = W^N(y_1^N | u_1^NG_N)$\end{center}

for appropriately defined alphabets on either side.

We shall talk about the implementations of this transformation and the encoding complexity later in this article.
\item \textbf{Channel splitting}

As mentioned earlier, we need to split $W_N$ back into a set of N channels. We call them virtual channels and denote them individually as $W_N^{(i)} : \chi \longrightarrow \Upsilon^N  \chi^{i-1}$, $1\leq i \leq N$ and define them as :

$W_N^{(i)}(y_1^N,u_i^{i-1}|u_i) \triangleq \underset{u_{i+1}^N \in \chi^{N-i}}{\sum}\frac1{2^{N-1}}W_N(y_1^N|u_1^N)$

This definition goes hand in hand with the successive cancellation decoder used in polar coding as we describe below. Let's say that we are trying to decode the i-th bit and we are given the correctly decoded estimates for the first i-1 bits. We use this vector of the i-1 estimates and the vector of observations $y_1^N$ to get the i-th bit. We assume that the inputs $u_1^N$ are uniformly distributed. Note that even if we tranform this vector into different vectors during the encoding process, the fact that the bits are uniform still holds because the transformations are permutations and are linear too. This uniformity dictates a factor of $\frac1{2^{N-1}}$ in the term.

\end{itemize}

Let us observe a few properties of the virtual channels we get after the above transformation. To make things easy, we shall see what happens when the channel is a BEC($\epsilon$) with uniform input. Since we are talking about uniform input, the capacity is $I(W)$.

\begin{center} $I(W_N^{(2i-1)}) = I(W_\frac N2 ^{(i)})^2$

$I(W_N^{(2i)}) = 2I(W_\frac N2 ^{(i)}) - I(W_\frac N2 ^{(i)})^2$

$I(W_1^{(1)}) = 1-\epsilon$
\end{center}

Observe that at every level of recursion, we get one channel which has capacity\footnote{the capacity of these channels is I(W) because the channel has uniform inputs} better than the original one and one which has capacity worse than the original one.\\

In order to analyse what happens for the general channel, we would like to see what happens locally to the above quantities. For doing this we would like to have the transtition probabilites of each of the virtual channels to be related to the original individual channels directly, instead of being related block by block. We want to map the independent copies of channel W to the channels we get after splitting i.e. $(W, W) \longrightarrow (W', W'')$

In more general terms, in the following proposition we map $(W_N^{(i)}, W_N^{(i)}) \longrightarrow (W_{2N}^{(2i-1)},W_{2N}^{(2i)}).$

\begin{prop}{Recursive channel transformations}

For any $n\geq 0, N = 2^n, 1\leq i \leq N$,
\center
$W_{2N}^{(2i-1)}(y_1^{2N},u_1^{2i-2}|u_{2i-1}) = \underset{u_{2i}}{\sum}\frac12W_N^{(i)}(y_1^N,u_{1,0}^{2i-2}\oplus u_{1,e}^{2i-2}|u_{2i-2}\oplus u_{2i})* W_N^{(i)}(y_{N+1}^{2N},u_{1,e}^{2i-2}|u_{2i})$

and

$W_{2N}^{(2i)}(y_1^{2N},u_1^{2i-1}|u_{2i}) = \underset{u_{2i}}{\sum}\frac12W_N^{(i)}(y_1^N,u_{1,0}^{2i-2}\oplus u_{1,e}^{2i-2}|u_{2i-2}\oplus u_{2i})* W_N^{(i)}(y_{N+1}^{2N},u_{1,e}^{2i-2}|u_{2i})$

\end{prop}

We are ready to talk about how the capacity and the reliability\footnote{The Bhattacharya parameter} change through a local transformation as above.

As we have seen in the erasure channel, we get one channel which is good and another which is bad\footnote{In loose terms good channels have more capacity and lesser reliability than the original channel and the opposite is true for bad channels}. the following proposition formalizes this notation

\begin{prop}Local transformation of rate and reliability

If $(W, W) \longrightarrow (W', W'')$ is the local transformation, then the following statements are true
\begin{enumerate}
\item $I(W') + I(W'') = 2I(W)$

\item $I(W') \leq I(W'')$

\item $Z(W'') = Z(W)^2$

\item $Z(W') \leq 2Z(W)-Z(W)^2$
\end{enumerate}
\end{prop}

From the above proposition, we can see that $I(W') \leq I(W) \leq I(W'')$ and $Z(W') \geq Z(W) \geq Z(W'')$, which goes with our intuition that we have one good and one bad channel. Also $Z(W')+Z(W'') \leq 2Z(W)$, which means that the reliability parameter can only improve with our transformation. Note that equality for the above is when we have perfect or useless channels. 

We apply these results to the recursive formulations in the previous proposition. We sketch the recursive transformations as a binary tree, in which every node gives birth to a good and a bad channel. The root is the original channel W and the channel $W_{2^n}^{(i)}$ is located at the n-th level and i-th node from top. We can label the nodes in this tree in a natural way - one in which each node is labelled with the path taken from the root ( 1 means up and 0 means down).

If we do this operation many number of times, we expect that most of the obtained channels have capacities near zero or one. In other words, the channel becomes significantly polarized after a few iterations. The following theorem formalizes our intuition.

\begin{thm} For binary channel W, the channels $\{W_N^{(i)}\}$ polarize in the sense that, for any fixed $\delta in (o,1)$  as N tends to infinity through powers of two, the fraction of indices for which $I(W_N^{(i)} \in (1-\delta, 1]$ goes to $1-I(W)$ and the fraction of those for which capacity lies in $[0, \delta)$ goes to $1-I(W)$.

Proof : \normalfont

The sequence of random variables $I_ n$ defined as the capacity of the channel obtained by starting at the node and taking the path n, is a martingale, because it is memoryless and $E[I_{n+1} | \text{path n}] = \frac 12 I(W_{path n, 0})+ \frac12 I(W_{path n,1}) = I_n$. Also, the sequence of random variables $Z_n$  is a supermartingale because  it is memoryless and $E[|Z_{n+1} | path n|] = \frac 12 Z(W_{path n 0})+ \frac12 Z(W_{path n1}) \leq Z_n$. Both the martingales are uniformly integrable and hence converge, by the martingale convergence theorem.

It follows that $E[Z_{n+1} - Z_n] \longrightarrow 0$ as $n\longrightarrow \infty$. Since $Z_{n+1} =Z_n^2$ with probability $ \frac12$, $E[|Z_{n+1} -Z_n|] \geq E[Z_n(Z_{n+1} -Z_n)] \geq 0$. By sandwich theorem of limits, $E[Z_n(1-Z_n)] \longrightarrow 0$ which implies that $E[ Z_\infty(1-Z_\infty)] = 0$. Hence $Z_\infty = 0 $ or $Z_\infty = 1 $ almost everywhere.

 Proposition 1 implies that $I_\infty$ takes values in $\{0,1\}$ with $P(I_\infty=1) = I_\infty$ and $P(I_\infty=0) =1- I_\infty$.
\end{thm}

We can thus see that polar codes indeed achieve capacity in the limit of the blocklength going to infinity.
\newpage
\subsection*{Rate of Polarization}
\addcontentsline{toc}{subsection}{Rate of Polarization}
Until now we have said that ploar codes with sufficiently high blocklengths acheives capacity. We would to know how fast this happens and the error probability for a particular block length. This section tries to address the above questions.

This question was answered by Arikan in his paper and is outlined in the following theorem. Guruswami improved upon the result in his paper and the result shall be mentioned later.
\begin{thm} 

For a BDMC W with $I(W) > 0$, and any fixed $R < I(W)$, there exists a sequence of sets $A_N \subset \{1, 2, 3, \dots,N\}$, $N = 2^n$ such that $|A_N|\geq NR$ and $Z(W_N^{(i)}) \leq O(N^{-\frac54})$ for all $i \in A_N$

\normalfont
\end{thm}

The above theorem essentially says that there exists a subset of the set of virtual channels which are 'good' and capacity does not decrease. This process also involves that the reliability factor goes down exponentially in the blocklength.

\subsection*{Polar Coding}
\addcontentsline{toc}{subsection}{Polar Coding}
We have seen in the earlier sections that the synthetic channels are sufficiently polarized. We need a way to access the 'good' channels - channels $W_N^{(i)}$ for which $Z(W_N^{(i)})=0$  and thus achieve the symmetric channel capacity. 

We define a class of codes called $G_N-coset$ $codes$ in which $G_N$ is the generator matrix i.e. $x_1^N = u_1^N G_N$. For an arbitrary subset A of the indices, we can write $x_1^N$ as $X_1^N = u_A G_N(A) \oplus u_{A^c}G_N(A^c)$ because it is a linear transformation. We have three parameters here - A, $u_A$ and $u_{A^c}$ and hence we talk about $(N, K, A,u_{A^c})$ codes. A is interpreted  as the 'information set', the set of indices which coincide with 'good' channels and $A^c$ as the set of 'bad' channels. $u_{A^c}$ are the '$frozen$ $bits$' and we leave $u_A$ to be free variables. We need to give a rule for selecting the information set A. As we shall see later, the way we choose the frozen bits does not have any effect on how well the coding scheme performs over symmetric channels.

We shall briefly talk about the decoder for polar codes, because that will give us insights on how we could possibly choose the information set and the frozen bits.

\subsection*{The Successive Cancellation Decoder}
\addcontentsline{toc}{subsection}{Successive Cancellation Decoder}
We shall be considering a (N,K,A,$u_{A^c}$) $G_N$-coset code in which $u_1^N$ has been encoded into a codeword $x_1^N$ and sent over the channel $W^N$. The decoder's task is to generate an estimate $\hat u_1^N$ of $u_1^N$, given the knowledge of A, $u_{A^c}$ and the channel output $y_1^N$. An obvious way to decode the $A^c$ bits is to set $\hat u_{A^c}$ = $ u_{A^c}$. We need a way to decode $u_{A^c}$. We do this by exploiting the structure of polar codes, a way which uses the bits we have already decoded and which treats the bits which are not decoded until now, as noise. We call this a successive cancellation decoder. Let's formalize this :\\
 
Our SC decoder outputs decisions $\hat u_i$ in order from i=1 to n such that,

 \begin{center}$\hat u_i \triangleq \begin{cases}
u_i, \hspace{19mm} \text{if } i \in A^c\\
h_i(y_1^N, \hat u_i ^{i-1}),\hspace{2mm} \text{if}\hspace{1mm} i \in A
\end{cases}$
\end{center}

where,\\
\begin{center}
$h_i(y_1^N, \hat u_i ^{i-1}) \triangleq \begin{cases}
0,\hspace{17mm} \text{if} \hspace{2mm} \frac{W_N^{(i)}(y_i^N, \hat u_1^{i-1}|0)}{W_N^{(i)}(y_i^N, \hat u_1^{i-1}|1)}\geq 1\\

1,\hspace{17mm} \text{otherwise}
\end{cases}$
\end{center}

These functions are similar to the ML decoding functions, but differ in that they assume the bits which we have not seen yet as noise, in other words as RVs.

We need to analyse the probability of error in this SCD framework. The error probabilities are denoted in a natural way.

\begin{defn} The probability of block error for a (N,K,A,$u_{A^c}$) assuming that each vector $u_A$ is sent uniformly is 
\center 
$P_e(N,K,A,u_{A^c}) \triangleq \underset{u_A \in \chi^K}{\sum}\frac 1{2^K} \underset{y_i^N\in \Upsilon^N : \hat u_1^N (y_1^N) \neq u_1^N}{\sum} W_N(y_1^N|u_1^N)$

\end{defn}
We also denote the average of the above error probability over all choices of $u_{A^c}$ by $P_e( N,K,A)$. We claim that the reliability still stays an upper bound on this error probability.

\begin{prop} 
$P_e(N,K,A) \leq \underset{i \in A}{\sum}Z(W_N^{(i)})$
\end{prop}

\begin{thm} The average probability of block error for polar coding under SC decoding goes down exponentially as $O(N^{-\frac14})$ for any BDMC W and a fixed rate lesser than the capacity.

\begin{center}$P_e(N,R) = O(N^{-\frac14})$\end{center}

\normalfont The proof easily follows from theorem 4 and the relation between block and bit error for SC decoder.
\normalfont

\end{thm}

Note that the above can be viewed as an existential result, in the sense that there exists a way of setting the frozen bits so that the error probability goes down exponentially. We have stronger results when the channel is symmetric. Let's observe a few properties of symmetric channels.

\begin{prop}
If a BDMC W is symmetric, then $W^N$, $W_N$ and $W_N^{(i)}$ are also symmetric.

\end{prop}
 The symmetries of the channel dictate proposition 4 is true for any way of setting the frozen bits.

\begin{thm} The probability of block error for polar coding under SC decoding goes down exponentially as $O(N^{-\frac14})$ for any symmetric BDMC W and a fixed rate lesser than the capacity, for $u_{A^c}$ fixed arbitrarily.
\begin{center}$P_e(N,K,A,u_{A^c}) = O(N^{-\frac 14})$\end{center}

\normalfont The idea used in the proof is that the events - making an error in the block and choosing the vector of frozen bits are independent events and thus we can freeze the bits ( to say $0_1^N$) without affecting the probability of error.
\end{thm}

\subsection*{Encoding and Decoding Complexity}
\addcontentsline{toc}{subsection}{Encoding and Decoding Complexity}
Polar codes are quite remarkable in the sense that both coding and decoding are polynomial in the block length; to be more precise, both take $O(N\log N)$ time steps on a sequential machine. It would be good to note that the structure of the encoding matrix can help us do it faster than the above on a parallel machine( $O(\log N )$ time).\\

As for the encoding compexity, $G_N$, which is an involutory permutation matrix can be written in terms of tensor products and we can also exploit its relation to Fast Fourier Transforms. The O(N log N) time obtained is due to the bit-indexing methods frequently used in FFTs.\\

It can be easily observed that the decoding complexity is $O(N \log N)$ because we check the ML like function recursively using log N many levels, each taking O(N) time.\\

We can thus see that polar coding is not just capacity achieving, but also something which is quite implementable in practice owing to the low encoding and decoding complexities.

\newpage
\section*{Reed Muller Codes}
\addcontentsline{toc}{section}{Reed Muller Codes}
Reed-Muller codes are one of the oldest families of error correcting codes and use concepts from algebra for the encoding and decoding process. The idea is to look at the message as the coefficients of a polynomial in a suitable degree and pass the suitable 

\begin{defn}{Reed Solomon Codes}

$RS_{F, S, n, k} (m) = f(\alpha_1), f(\alpha_2), f(\alpha_3), \dots, f(\alpha_n))$, where $f(X) = m_0 +m_1 X + \dots + m_k X^k$.

\normalfont We view a message of k symbols as the coefficients of a univariate polynomial f(X) of degree k-1. We encode the message as the evaluations of this polynomial at n different points in the underlying field (or in a subset S which the code designer is left to choose).
\end{defn}

We should talk about how we are getting these evaluations across. We define a special matrix to make representation easier to work with.
\begin{defn} {The Vandermonde matrix}

G =
${\begin{bmatrix}
1 & 1 &1 & \dots &1\\
\alpha_1 & \alpha_2 &\alpha_3& \hdots &\alpha_n\\
\alpha_1^2 &\alpha_2^2 &\alpha_3^2 & \hdots & \alpha_n^2 \\
\vdots &\vdots & \vdots&\ddots &\vdots\\
\alpha_1^{k-1} & \alpha_2^{k-1} &\alpha_3^{k-1}& \dots &\alpha_n^{k-1}\\
\end{bmatrix}}$\\
 is the generator matrix for $RS_{F,s,n,k}$\\
\end{defn}

Looking at the generator matrix we can see that RS codes are linear.

\begin{prop}RS codes are linear.
\end{prop}

\begin{prop} The minimum distance of RS codes is (n-k+1).

Proof : \normalfont This is true because if $m' \neq m''$ are the messages, the corresponding polynomials have to differ in more than n-k locations. 
\end{prop}

RS codes are good in the sense that their distance is huge, but on the downside they require that the underlying field should be sufficiently large - at least of order n. To address this difficulty, we talk about Reed Muller codes. They are generalizations of Reed Solomon codes in the sense that we would be using multivariate polynomials instead.
\begin{defn} Reed Muller Codes

Given a field size q and a number m of variables, and a total degree bound r, the RM$_{q,m,r}$ code is the linear code over $F_q$ defined by the encoding map 

$f( X_1, X_2, \dots X_m)\longrightarrow <f(\alpha)> |_{\alpha \in F_q^m}$

applies to the domain of all polynomials in $F_q[ X_1, X_2, \dots X_m]$ of total degree def(f)$\leq r$
\end{defn}

Let's talk about decoding these families of codes. We would expect that unique decoding is possible only if there are not too many errors in the code.
\begin{thm}{Unique decoding is possible only if the distance of the code is at least $\frac{n-k}2$}.

Proof: \normalfont We look at hamming balls and for the boundary condition, we want that the point is close to exactly two of them, which gives us the result.
\end{thm}

RS codes are decoded using the 'magical' Berlekamp-Welch algorithm which involves fitting the bad points in a curve and then finding them out.

Let $y_i$s  be the evaluations of the polynomial at distinct locations $x_i$ for $i \in \{1,2, \dots n\}$. Let e be the number of errors. Our obective is to find a polynomial p(X) of degree at most k-1 such that the number of errors e is respected. The following algorithm helps us in doing so.
\begin{alg}Berlekamp-Welch algorithm
\begin{enumerate}
\item If there is a polynomial such that $p(x_i) = y_i$ for all $i= 1,2 \dots n$, then output p. Otherwise :

\item Find polynomials E(x) and N(x) such that :\begin{itemize}
\item E is not identically zero
\item E(x) has degree at most e and N(x) has degree at most e+k-1
\item For every $i =1, 2, \dots n$, $N(x_i)= E(x_i)*y_i$
\end{itemize}
\item Output $\frac{N(x)}{E(x)}$ if E(x) divides N(x), else output error.
\end{enumerate}
\end{alg}
We write the constraints for the polynomials E(x) and N(x) and then find them using the above algorithm. It can be proved that any solution for the constraints satisfies the conditions and gives us the correct generator polynomial.

If unique decoding is not possible, then we can do list decoding till a particular fraction of errors after which the list size becomes exponential in size. This is done using the even more magical Guruswami-Sudan algorithm, which uses the same idea of fitting the bad points in a curve and then finding them out.

\begin{defn}{1-transitive Codes}

A code $\mathscr C$ is said to be 1-transitive if for any $j_1$ and $j_2$ $\in$ [N] satisfying  $j_1 \neq j_2$, there exists a permutation $\pi$ : [N] $\rightarrow$ [N] such that :  
\begin{itemize}
\item $\pi(j_1) = j_2$ 
\item $y _{\pi(1)},y _{\pi(2)} . . y _{\pi(n)} \in \mathscr C$ for every  $y _{1},y _{2} . . y _{n} \in \mathscr C$
\end{itemize}
\end{defn}
\newpage
\begin{defn}{2-transitive Codes}

A code $\mathscr C$ is said to be 2-transitive if for any $j_1$, $j_2$,$j_3$, $j_4$ $\in$ [N] satisfying  $j_1 \neq j_2$ and $j_3 \neq j_4$ , there exists a permutation $\pi$ : [N] $\rightarrow$ [N] such that :  
\begin{itemize}
\item $\pi(j_1) = j_3$
\item $\pi(j_2) = j_4$ 
\item $y _{\pi(1)},y _{\pi(2)} . . y _{\pi(n)} \in \mathscr C$ for every  $y _{1},y _{2} . . y _{n} \in \mathscr C$

\end{itemize}
\end{defn}

Let's look at some important properties of the above codes.
\begin{prop}{Reed Solomon codes are 2-transitive}

Proof : \normalfont As we have seen before, RS codes are  generated by the Vandermonde matrix $G$. The input $m_1^N$ is transformed to the codeword $y_1^N$ using the matrix $G$, which is eventually sent over the channel. 

Formally, $y_1^N = m_1^N * G$

We are given four locations in the code  - say a, b, c and d $\in [N]$ such that $ a \neq b$ and $c\neq d$ and we need to give a permutation $\pi : [N] \longrightarrow [N]$  such that $\pi(a) =c$ and $\pi( b)=d$ and also preserves membership in the code.

For the moment, pick any permutation which satisfies the above constraints - we shall see why it does not matter. Observe that this permutation is a $N \times N$ 0,1 permutation matrix.

$ \underset{ N\times N}{\pi} * \underset{1 \times N}{\begin{bmatrix}m_1&m_2&m_3& \hdots & \hdots& m_n \end{bmatrix}} = \begin{bmatrix}m_{\pi(1)}&m_{\pi(2)}&m_{\pi(3)}& \hdots & \hdots& m_{\pi(n)} \end{bmatrix} $

Same holds for the vector $y_1^N$, because the $\pi$ matrix does not care about the actual values of the vector it is multiplied with (for it is a permutation matrix). It only permutes the indices to get another vector and will do the same for $m_1^N$. 

i.e.

$ \underset{ N\times N}{\pi} * \underset{1 \times N}{\begin{bmatrix}y_1&y_2&y_3& \hdots & \hdots& y_n \end{bmatrix}} =\underset{1 \times N}{ \begin{bmatrix}y_{\pi(1)}&y_{\pi(2)}&y_{\pi(3)}& \hdots & \hdots& y_{\pi(n)} \end{bmatrix}}$

We are given $y_1^N = m_1^N * G$

Premultiply the encoding equation with the matrix $\pi$

\hspace{2mm}$\pi * \begin{bmatrix}y_1&y_2&y_3& \hdots & \hdots& y_n \end{bmatrix}=\pi * \begin{bmatrix}m_1&m_2&m_3& \hdots & m_n\end{bmatrix} * G$

$\implies  \begin{bmatrix}y_{\pi(1)}&y_{\pi(2)}&y_{\pi(3)}& \hdots & \hdots& y_{\pi(n)} \end{bmatrix}= \begin{bmatrix}m_{\pi(1)}&m_{\pi(2)}&m_{\pi(3)}& \hdots & \hdots& m_{\pi(n)} \end{bmatrix}* G$

Since RS codes are linear codes, and we are giving evalutions over the whole field,   $m_{{\pi(1),\pi(2),\pi(3)} \dots \pi(n)}$ is a codeword too.

\end{prop}
\begin{prop}{Reed Muller codes are 2-transitive}

Proof is similar to the one given above, find any transformation which satisfies our constraints works.
\end{prop}
Note that if we want our codes to satisfy these transitive properties, we must give evaluations of the polynomials on all points in the corresponding fields.\\

We shall see in the coming sections that RM codes have more interesting properties.
\newpage
\section*{Reed Muller Codes Achieve Capacity on Erasure Channels}
\addcontentsline{toc}{section}{Reed Muller Codes on Erasure Channels}
As recently as July 2015, Urbanke et. al. and Santhosh Kumar et. al. independently proved that RM codes achieve capacity on erasure channels. Let's us now build up the machinery required to prove the aforementioned result. We assume that the ith bit of the message is transmitted through an erasure channel with probability $p_i$ i.e. BEC$(p_i)$. Let's denote this channel by BEC(\underline p)\footnote{ \underline p is the corresponding vector of erasure probabilities}. In this setting we would like to analyze the probability that the MAP decoder is unable to decode the i-th bit and then try to get a bound on the probability of error for the block MAP decoder. We also assume that the input distribution is uniform. \\

Since we are talking about erasure channels and MAP decoding, we should define EXIT functions which we would use later to capture the decoding errors.
\begin{defn}{EXIT functions}

The vector EXIT function associated with the ith bit is defined to be
\center $h_i(\underline p) \triangleq H (X_i | \underline Y_{\sim i}(\underline p _{\sim i})$ 
\end{defn}
We can define an average exit function, which we would use later.
\begin{defn}{Average EXIT function}

Average EXIT function is defined as
\center $h(\underline p) \triangleq \sum ^ N _ {i=1}h_i(\underline p) $
\end{defn}
We denote the bit MAP decoder's output for ith bit as D$_i$. On receiving the sequence \underline Y, if the ith bit $X_i$ can be recovered uniquely, then D$_i($\underline Y) = X$_i$. Otherwise, D$_i$ declares an erasure and returns *. We claim that the probability of the decoding failing to decode is equal to the ith EXIT function.
\begin{prop}{Pr ( $D_i (\underline Y) \neq X_i) = H( X_i | \underline Y)$ }

Proof : \normalfont 

Whenever bit i can be recovered from a received sequence \underline Y = \underline y, $ H( X_i \underline Y = \underline y) = 0$. Otherwise, $ H( X_i \underline Y = \underline y) = 1 $ because of the uniform codeword assumption.
\end{prop}
Observe that the probability of being able to decode the ith bit, and hence the value of the ith EXIT function is either equal to zero or one; it cannot be anything between zero or one.

\begin{prop}{The MAP exit function for the ith bit satisfies $h_i(\underline p) = \frac{\partial H(\underline X | \underline Y(\underline p))}{\partial p_i}$}

Proof : \normalfont 

By chain rule of entropy $H(\underline X | \underline Y(\underline p))= H(X_i |\underline Y(\underline p)) + H(X_{\sim i} |X_i,\underline Y(\underline p))$

Observe that the second term written in the above expansion is independent of $p_i$ because $ H(X_{\sim i} |X_i,\underline Y(\underline p)) =  H(X_{\sim i} |X_i,\underline Y_{\sim i}(\underline p_{\sim i}))$

$H(\underline X | \underline Y(\underline p))=Pr(Y_i = *) H(X_i |\underline Y_{\sim i}(\underline p_{\sim i}, Y_i = * )+ Pr(Y_i = X_i)H(X_i |\underline Y_{\sim i}(\underline p_{\sim i}, Y_i = X_i ) \\= p_i H(X_i |\underline Y_{\sim i}(\underline p_{\sim i}))$

The second entropy term is zero and the proposition thus follows.
\end{prop}
The above proposition leads us to an important theorem in coding theory - the area theorem. 
\begin{thm}{The Area Theorem} 

The average EXIT function satisfies the area theorem : \\
\begin{center}$\int _0^1 h(p)dp = \frac KN$.\end{center}

Proof : \normalfont

The above proposition gives us the derivative of the function. It follows that we can integrate from 0 to 1 on a fixed path $(p, p, p, \dots p)$.

$H(\underline X | \underline Y(\underline p)(1)) -H(\underline X | \underline Y(\underline p)(0)) = \int_0^1 (\sum_{1=1}^N h_i(t))dt$

$H(\underline X | \underline Y(\underline p)(1))=$H(\underline X ) = K since it is uniform distribution and the encodings capture the same randomness as that of the original distribution. 

$H(\underline X | \underline Y(\underline p)(0)) = 0$.

Hence the result.

\end{thm}

We would like to look at the set of erasure patterns ( vectors $Y_{\sim i}$ from which we cannot decode the ith bit $X_i$ indirectly using MAP decoding. We claim that the following set correctly captures this notion and contains all the erasure patterns from which indirect recovery of the ith bit is not possible. Hence the measure of this set determines the probability of error which is what we eventually want.

\begin{defn} {The set of patterns 'bad' for bit i are contained in $\Omega_i$ defined as }

\center$\Omega_i \triangleq \{ A \subseteq [N] \setminus {i} | \exists B \subseteq [N] \setminus {i}, B \cup \{i\} \in \mathscr C, B \subseteq A \}$

\end{defn}

From the above discussion, it is clear that the measure of the set is equal to the probability of error, which, in turn is equal to the ith exit function because of the uniform input assumption. We summarize this in the following proposition.

\begin{prop}{$\Omega_i$ encodes $h_i(\underline p)$}

\center $h_i (\underline p) = \mu _{\underline p}( \Omega_i) = \underset {A \in \Omega_i}{ \sum}\underset{l \in A}{\Pi}p_l\underset{l \in A^c \setminus \{i\}}{\Pi} (1 - p_l)$

\end{prop}

We claim that if the code $\mathscr C$ is 2-transitive, the set $\Omega_i$ is 1-transitive.
\begin{prop}{If C is 2-transitive then $\Omega_i$ is 1-transitive }

Proof : \normalfont

Since the code is 2 transitive, there exists a permutation $\pi$ such that $\pi(i)=j$ for all $i\neq j$. We need to show that this permutation preserves membership in the corresponding $\Omega s$. In other words, we need to show that if A $\in \Omega_i$ then $\pi(A) \in \Omega_{\pi(i)}$.

Since $A\in \Omega_i$, $\exists B \subseteq A$ such that $B \cup {i} \in \mathscr C$. $\pi(B\cup{i}) \in \mathscr C$. Observe that  $\pi(B\cup{i}) = \pi (B) \cup \pi(i) =  \pi (B) \cup j$. Since $\pi(B) \subseteq \pi(A)$, it follows that $\pi(A) \in \Omega_j$
\end{prop}
This is a bijection because we can do what we did above with the indices i and j interchanged.
\begin{prop}{All EXIT functions are equal.}

Proof:
\normalfont
Since the code is transitive, any two locations have a permutation between them. The corresponding $\Omega s$ have a bijection between them, and the EXIT function is equal to the measure of the corresponding set $\Omega $. The proposition thus follows. 
\end{prop}
Note that all EXIT functions are equal to the average EXIT function and thus we are free to invoke the area theorem now.\\

Intuitively, if we have an erasure pattern which is 'bad', we will not be able to decode the patterns which are obtained after adding more erasures at places where there were no erasures before. The following propositon formalizes this.
\begin{prop}{$\Omega_i $ is monotone}

Proof :\normalfont

We have to prove that if $A \in \Omega_i$ and $A\subseteq C$  then such that $C \in \Omega_i$

Looking at the definition of $\Omega_i$ there exists $ B \in [N]\setminus \{i\}$ such that $ B \subseteq A,$ $B \cup \{i\} \in \mathscr C,$ and $ B \subseteq A $. $B\subseteq A \subseteq C$ and hence it follows that $C \in \Omega_i$
\end{prop}

Let's qiuckly look at what we know until now. \begin{enumerate}
\item  $h_i(p)$ captures the probability of error of MAP decoder 
\item $\Omega_i$ encodes $h_i(p)$
\item All EXIT functions are equal to the average exist function h.
\item The area under the h vs p curve is the rate ( using area theorem)
\end{enumerate}

If we prove that the set  $\Omega_i$ has a sharp threshold, then we would've proved that 2-transitive codes achieve capacity, since the threshold would occure at p = 1-R.\\

We define another set, which would be useful to prove the sharp threshold behaviour - the set of erasure patterns for which location j is pivotal in the indirect recovery of the i-th bit. In other words, flipping the j-th bit flips the erasure pattern between $\Omega_i$ and $\Omega_i^c$.

\begin{defn}{The set of erasure patterns for which the j-th bit is pivotal in the indirect decoding of the i-th bit is }

$\partial_j \Omega _ i \triangleq \{ A \subseteq [N]\setminus \{i\}|A\setminus \{j\}  \not \in \Omega_i, A\cup \{j\} \in \Omega_ i\}$
\end{defn}
Note that $\partial_j \Omega _ i$ contains patterns from both $\Omega_i$ and $\Omega_i^c$.

Intuitively we expect that once we permute the locations to another set of locations, the bits which were pivotal before, stay pivotal in the permuted world, for if this were not to happen, we could have magically decoded the concerned bit using this permutation. We formalize this notion.

\begin{prop}{If a code $\mathscr C $ is 2-transitive, then for distinct i, j, k $\in [N]$, there exists a bijection between $\partial_j \Omega _ i$ and $\partial_k \Omega _ i$.}

Proof : \normalfont

Consider a codeword A. Since the code is 2-transitive, there exists a permutation $\pi$ such that $\pi(i) = i$ and $\pi(j) = k$ and $\pi(A) \in \mathscr C$. We need to prove that if A $\in$ $\partial_j \Omega _ i$ then $\pi(A)$ $\in$ $\partial_k \Omega _ i$. 
\begin{itemize}
\item 
Case 1 : $A \cup \{j\} \in \Omega_i$ and $A \setminus \{j\} \not \in \Omega_i$.\\
 $\pi(A) \in \Omega_i$ and $\pi(A\setminus\{j\}) \not \in \Omega_i$ because of the transitivity of $\Omega_i$.\\
$\pi(A\setminus \{j\}) = \pi(A) \setminus \{k\}$. Thus $\pi(A) \in \partial_k \Omega _ i$

\item Case 2 : $A \cup \{j\}\in \Omega_i$ and $A \not \in \Omega_i$.

\end{itemize}

If we were to interchange the indices j and k, a similar argument would hold. Hence there is a one-to-one correspondence between the two sets.

\end{prop}
The measures of the sets $\partial_j \Omega _ i$ and $\partial_k \Omega _ i$ are equal since we prove that there is one to one correspondence between them. 

$\mu_{\underline p} ( \partial_j \Omega _ i) =  \underset {A \in \partial_j \Omega_i}{ \sum}\underset{l \in A}{\Pi}p_l\underset{l \in A^c \setminus \{i\}}{\Pi} (1 - p_l)$

Our quest to prove that the set $\Omega$( and in turn the average EXIT function) has a sharp threshold, requires us to talk about the influences of the variables and invoke suitable results from boolean function analysis. Let's define a few terms first.

\begin{defn}{Influence of a variable}

\begin{normalfont} {Let $\Omega $ be a monotone set and let $\partial_j \Omega \triangleq \{ \underline x \in \{ 0,1\}^N | \mathbb{1}_\Omega (\underline x) \neq\mathbb{1}_\Omega (\underline x^{(j)})\}$}, where $\underline x^{(j)}$ is defined by $x_l^{(j)} = x_l$ for $l \neq j$  and $x_j^{(j)} =1 - x_j$

\end{normalfont}
\center The influence of bit j $\in$  [N] is defined by $I_j^{(p)}(\Omega) \triangleq \mu_p (\partial_j \Omega )$
\end{defn}

\begin{defn}{Total Influence}

The total influence of a bunch of variables is defined as
\center $I^{(p)} ( \Omega)  = \sum^N_{l=1} I _l^{(p)} ( \Omega)$
\end{defn}

Let's look at a result which talks about the derivative of the measure of monotone sets.
\begin{lem}{Margulis - Russo Lemma}

Let $\Omega$ be a monotone set, then

\center $\frac{d_{\mu_p(\Omega)}}{dp} = I^{(p)}(\Omega)$
\end{lem}
The above result says that the derivative is lower bounded by a quantity ( which as we shall see, scales with N). But the value of the function has to sum up to one since it is a graph which plots the measure against the probability. This implies that the derivative cannot be high everywhere, and hence the function should have a sharp threshold.

First we shall show how this is related to the problem at hand - how the influence of the EXIT functions is related to the measure of the set of pivotal bits.

\begin{prop} 
$\frac{\partial h_i(\underline p)}{\partial p_j} = \underset{\partial_j \Omega _ i}{\sum} \underset{l \in A}{\Pi}p_l \underset{l \in A^c\setminus \{i\}}{\Pi}(1-p_l )$

Proof : \normalfont We evaluate the partial derivative from the explicit evaluation of $ h_i(\underline p))$ done in Proposition 6.

$h_i (\underline p) = \mu _{\underline p}( \Omega_i) = \underset {A \in \Omega_i}{ \sum}\underset{l \in A}{\Pi}p_l\underset{l \in A^c \setminus \{i\}}{\Pi} (1 - p_l)$

If we differentiate the above quantity with respect to $p_j$ and use the fact that $\Omega_i$ is monotone, we get the above result.

\end{prop}
We still need to show how these influences scale with N and tie up the loose ends, and the following theorem, which we state without proof helps us in doing so.
\begin{thm}{}
Let $\Omega$ be a montone set and suppose that, for all $0\leq p \geq1$, the influences of all bits are equal $I_1^{(p)}(\Omega) = I_2^{(p)}(\Omega) = \dots =  I_N^{(p)}(\Omega)$. The following is true :
\begin{enumerate}
\item There exists an universal constant $C \geq 1$ which is independent of p, $\Omega$ and N, such that 

 $\frac{d_{\mu_p(\Omega)}}{dp} \geq C( \log N)( \mu_p(\Omega))(1-\mu_p(\Omega))$

\item  For any $0 < \epsilon \leq \frac12$,

$p_{1-\epsilon} - \ p_\epsilon \leq \frac2C \frac{\log \frac{1-\epsilon}{\epsilon}}{\log M}$

where $p_t \triangleq h^{-1} = inf \{ p \in [0,1] | h(p) \geq t\}$ is the inverse function for the average EXIT function\footnote{ h(p) is a strictly increasing continuous polynomial function and hence inverse is well-defined on [0,1]}.

\end{enumerate}

\end{thm}

It is important to note that for the above theorem to hold, the influences have to be spread quite uniformly. Let's see why this is intuitively true. Say there is a (dictator)function which has N variables and only one bit is influential ( output depends only on this particular bit), the total influence is p (the success). The derivative of this with respect to p is 1 and it does not scale with N as the RHS of the first expression is supposed to.\\

Observe that the second statement implies that the set has a sharp threshold because 

$p_{1-\epsilon} - \ p_\epsilon \longrightarrow 0$ as N $\longrightarrow \infty$\\

We have considered bit-MAP decoding for the above result. This result also implies that the channel is capacity achieving under block-MAP decoding, if we consider the following proposition which is quite elementary in itself.

\begin{prop}{ Relation between the error probabilities of the bit MAP and block MAP decoder}

\begin{center} $P_{block-MAP} \leq \frac{N P_{bit-MAP}}{d_{min}}$

\end{center}

where $d_{min}$ is the minimum distance of the code.
\end{prop}

We can see that if $P_{bit-MAP} \longrightarrow 0$ with sufficient speed, then $P_{block-MAP} \longrightarrow 0$ as well and thus can summarize the result in the following theorem.
\begin{thm}{ 2-transitive codes achieve capacity on erasure channel under MAP decoding.}

\end{thm}
\begin{cor}
Reed Muller codes achieve capacity on erasure channels.
\end{cor}

\begin{thm}{Polar codes are 2-transitive}

Proof : \normalfont The proof is exactly similar to the one given while proving that RS codes are 2-transitive. As we have seen before, polar codes are $G_N$-coset codes in which the input $u_1^N$ is transformed to the codeword $x_1^N$ using the matrix $G_N$, which is eventually sent over the channel. 

$x_1^N = u_1^N G_N$

We are given four locations in the code  - say a, b, c and d $\in [N]$ such that $ a \neq b$ and $c\neq d$ and we need to give a permutation $\pi : [N] \longrightarrow [N]$  such that $\pi(a) =c$ and $\pi( b)=d$ and also preserves membership in the code.

For the moment, pick any permutation which satisfies the above constraints - we shall see why it does not matter. Observe that this permutation is a $N \times N$ 0,1 permutation matrix.

$ \underset{ N\times N}{\pi} * \underset{1 \times N}{\begin{bmatrix}u_1&u_2&u_3& \hdots & \hdots& u_n \end{bmatrix}} = \begin{bmatrix}u_{\pi(1)}&u_{\pi(2)}&u_{\pi(3)}& \hdots & \hdots& u_{\pi(n)} \end{bmatrix} $

Same holds for the vector $x_1^N$, because the $\pi$ matrix does not care about the actual values of the vector it is multiplied with (for it is a permutation matrix). It only permutes the indices to get another vector and will do the same for $u_1^N$. 

i.e.

$ \underset{ N\times N}{\pi} * \underset{1 \times N}{\begin{bmatrix}x_1&x_2&x_3& \hdots & \hdots& x_n \end{bmatrix}} = \underset{1 \times N}{\begin{bmatrix}x_{\pi(1)}&x_{\pi(2)}&x_{\pi(3)}& \hdots & \hdots& x_{\pi(n)} \end{bmatrix}} $

Premultiply the polar encoding equation with the matrix $\pi$

\hspace{2mm}$\pi * \begin{bmatrix}x_1&x_2&x_3& \hdots & \hdots& x_n \end{bmatrix}=\pi *\begin{bmatrix}u_1&u_2&u_3& \hdots & \hdots& u_n \end{bmatrix} * G_N$
\end{thm}

$\implies$ $\begin{bmatrix}x_{\pi(1)}&x_{\pi(2)}&x_{\pi(3)}& \hdots &x_{\pi(n)}\end{bmatrix} = \begin{bmatrix}u_{\pi(1)}&u_{\pi(2)}&u_{\pi(3)}& \hdots &u_{\pi(n)}\end{bmatrix} * G_N$

Polar codes are linear codes and thus $u_{{\pi(1),\pi(2),\pi(3)} \dots \pi(n)}$ is a codeword, $x_{{\pi(1),\pi(2),\pi(3)} \dots \pi(n)}$ is a codeword too.
The above theorem, along with the results related to 2-transitive codes in this section gives us another proof of the fact that polar codes achieve capacity on the binary erasure channel. 

\begin{cor}
Polar codes acheive capacity on erasure channels under MAP decoding.
\end{cor}It should be noted that the rate of polarization in Guruswami's paper on \textit{'Speed of Polarization'} would be faster than the above proof would give us.
\newpage
\section* {Future Work}
\addcontentsline{toc}{section}{Future Work}
The next big question we would like to ask is whether Reed Muller codes achieve capacity for other symmetric channels. This approach of EXIT functions and monotone thresholds does not work even for the BSC. \\

Another interesting question would be whether we can improve the above results by coming up with some non trivial decoding scheme instead of the expensive MAP decoding.

\addcontentsline{toc}{section}{Bibliography}

\begin{thebibliography}{9}

\bibitem{arikan}
  Erdal Arikan,
  \emph{"Channel polarization: A method for constructing capacity-achieving codes for symmetric binary-input memoryless channels"},
  IEEE International Symposium on Information Theory, 2008.

\bibitem{guruswami}
Venkatesan Guruswami, Patrick Xia,
\emph{"Polar Codes: Speed of polarization and polynomial gap to capacity"},
IEEE Foundations of Computer Science (FOCS), 2013.

\bibitem{RM1}
Shrinivas Kudekar, Marco Mondelli, Eren Sasoglu, Rudiger Urbanke
\emph{"Reed-Muller Codes Achieve Capacity on the Binary Erasure Channel under MAP Decoding."},
arXiv:1505.0583, 2015.


\bibitem{RM2}
Santhosh Kumar and Henry D. Pfister,
\emph{"Reed-Muller Codes Achieve Capacity on Erasure Channels"},arXiv:1505.05123, 2015.

\bibitem{errorcorrection}
Venkatesan Guruswami, Atri Rudra,
\emph{"Error-correction up to the information-theoretic limit"}
Communications of the ACM, Volume 52 Issue 3, March 2009.

\bibitem{shannon}
C. E. Shannon,
\emph{"
A mathematical theory of communication"}
ACM SIGMOBILE Mobile Computing and Communications Review, Volume 5 Issue 1, January 2001.
\end{thebibliography}
\end{document}